\documentclass[letterpaper]{JHEP3}
\usepackage[T1]{fontenc}
\usepackage[latin9]{inputenc}
\usepackage{array}
\usepackage{graphicx}
\usepackage{nicefrac}
\usepackage{amsmath,epsfig}
\usepackage{amssymb,amsfonts,mathrsfs}
\usepackage{color}

\usepackage{setspace}

\usepackage{graphics}
\usepackage{epsfig}
\usepackage{amsfonts}

\newcommand{\Tr}{\mbox{Tr}}

\newcommand{\Q}{{\cal Q}}

\global \long \def \NN{ \mathcal{N}}
\global \long \def \II{ \mathcal{I}}

\def\a{\alpha}
\def\b{\beta}

\def\h{\eta}

\def\IC{\relax\hbox{$\inbar\kern-.3em{\rm C}$}}

\def\IC{{\bf C}}

\def\bea{\begin{eqnarray}}
\def\eea{\end{eqnarray}}
\def\be{\begin{equation}}
\def\ee{\end{equation}}
\def\ba{\begin{align}}
\def\ea{\end{align}}
\def\bse{\begin{subequations}}
\def\ese{\end{subequations}}
\def\1F1{{}_1\!F_1}
\def\2F0{{}_2\!F_0}

\def\a{\alpha}

\def\h3{$\textrm{H}_3^+$}

\def\IC{{\mathbb C}}

\def\Tr{{\rm Tr}}

\def\lbldef#1#2{\expandafter\gdef\csname #1\endcsname {#2}}

\def\href#1#2{#2}

\newcommand{\beq}{\begin{equation}}
\newcommand{\eeq}{\end{equation}}
\newcommand{\ber}{\begin{eqnarray}}
\newcommand{\eer}{\end{eqnarray}}

\def\be{\begin{eqnarray}}
\def\ee{\end{eqnarray}}

\def\({\left(}
\def\){\right)}
\def\[{\left[}
\def\]{\right]}
\def\<{\langle}
\def\>{\rangle}

\def\II{{\cal I}}

\title{Reducing the 4d Index to the $S^3$ Partition Function}
\preprint{YITP-SB-11-12}
\author{Abhijit Gadde\footnote{abhijit@insti.physics.sunysb.edu}$\;$ and
Wenbin Yan\footnote{wyan@insti.physics.sunysb.edu}
\\
\\
\it C.N. Yang Institute for Theoretical Physics,\\
\it Stony Brook University, \\
\it Stony Brook, NY 11794-3840, USA}

\bigskip

\abstract{
The superconformal index of a $4d$ gauge theory is computed by a matrix integral arising from localization of the  supersymmetric path integral on $S^3\times S^1$ to the saddle point.
As the radius of the circle goes to zero, it is natural to expect that the $4d$ path integral becomes the partition function of dimensionally reduced gauge theory on $S^3$. We show that this is indeed the case and recover the matrix integral of Kapustin, Willett and Yaakov from the matrix integral that computes the superconformal index.  Remarkably, the superconformal index of the ``parent'' $4d$ theory can  be thought of as the $q$-deformation of the $3d$ partition function.

}


\begin{document}
\section{Introduction}
String/M theory has led to a rich web of non-perturbative dualities between supersymmetric field theories. Checking/exploiting/extending these dualities requires exact computations in field theories. In recent years, using methods based on localization, several exact quantities in supersymmetric gauge theories have been computed. Two of such quantities, the superconformal index of $4d$ gauge theories \cite{Kinney:2005ej,Romelsberger:2005eg} and the partition function of supersymmetric gauge theories on $S^3$ \cite{Kapustin:2009kz,Kapustin:2010xq}, are the main focus of this note.

The superconformal index of ${\cal N}=1$ IR fixed points was computed in \cite{Dolan:2008qi, Spiridonov:2008zr, Spiridonov:2009za}, there it served as a check of Seiberg duality. The indices of ${\cal N}=4$ SYM and type IIB supergravity in $AdS_5$ were computed and matched in \cite{Kinney:2005ej}. The superconformal index of ${\cal N}=2$ supersymmetric gauge theories was used to check  ${\cal N}=2$ S-dualities conjectured by Gaiotto and to define a $2d$ topological field theory in the process \cite{Gadde:2009kb,Gadde:2010te}.
Recently the partition function of supersymmetric gauge theories on $S^3$ has been used to check a variety of $3d$ dualities including mirror symmetry \cite{Kapustin:2010xq} and Seiberg-like dualities \cite{Kapustin:2010mh}. Remarkably, the exact partition function has also allowed for a direct field theory computation of $N^{3/2}$ degrees of freedom of ABJM theory \cite{Drukker:2010nc,Herzog:2010hf}. The $S^3$ partition function of ${\cal N}=2$ theories is extremized by the exact superconformal R-symmetry \cite{Jafferis:2010un,Martelli:2011qj,Jafferis:2011zi} so just like the $a$-maximization in $4d$, the $3d$ partition function can be used to determine the exact R-charges at interacting fixed points. The purpose of this note is to relate these two interesting and useful exactly calculable quantities in $3$ and $4$ dimensions.

The superconformal index of a $4d$ gauge theory can be computed as a path integral on $S^3 \times S^1$ with supersymmetric boundary
conditions along $S^1$. All the modes on the $S^1$ contribute to this path integral. In a limit with the radius of the circle shrinking to zero
 the higher modes  become very heavy and decouple. The index is then given by a path integral over just the constant modes on the circle.
 In other words, the superconformal index of the $4d$ theory reduces to a partition function of the dimensionally reduced $3d$ gauge theory
on $S^3$. The $3d$ theory preserves all the supersymmetries of the ``parent'' $4d$ theory on $S^3 \times S^1$.

More generally, for any $d$ dimensional manifold $M^d$, one would expect the index of a supersymmetric theory on $M^d\times S^1$ to reduce to the exact partition function of dimensionally reduced theory on $M^d$. This idea was applied by Nekrasov to obtain the partition function of $4d$ gauge theory on $\Omega$-deformed background as a limit of the index of a $5d$ gauge theory~\cite{Nekrasov:2002qd}.

A crucial property of the four dimensional index that facilitates its computation is the fact that it can be computed exactly by a saddle point integral. We show that in the limit of vanishing circle radius, this matrix integral reduces to the one that computes the partition function of $3d$ gauge theories on $S^3$ ~\cite{Kapustin:2009kz,Kapustin:2010xq}. It doesn't come as a surprise as the path integral of the ${\cal N}=2$ supersymmetric gauge theory on $S^3$ was also shown to localize on saddle points of the action.

The note is organized a follows. In section \ref{pathint} we write the superconformal index of $4d$ theory as a saddle point integral and describe the limit in which this integral reduces to the $S^3$ partition function. The limit is performed in section \ref{mmlimit}. In particular, we show that the building blocks of the matrix model that computes the superconformal index in $4d$ map separately to the building blocks of the $3d$ partition function matrix model. In section \ref{dualities}, we comment on the connections between $4d$ and $3d$ dualities.
We conclude with an appendix that generalizes the Kapustin et. al. matrix model for ${\cal N}=4$ gauge theories with two supersymmetric deformations. One such deformation involving squashed $S^3$ was studied in \cite{Hama:2011ea}.

\

\noindent{\bf Note added}: While this note was in preparation we received \cite{Dolan:2011rp} in which the authors find a mathematical relation
between the 4d index and the $S^3$ partition function which is equivalent to the relation that we derive here from physical considerations.

\section{\label{pathint}$4d$ Index as a path integral on $S^3\times S^1$}

The superconformal index is a Witten index with respect to one of the supercharges.
 For concreteness, let us restrict ourselves to the supercharge \footnote{The supercharges of $\NN=2$ gauge theory are denoted as $\Q^I_\alpha$ and $\bar\Q_{I\dot{\alpha}}$ where $I=1,2$ is an $SU(2)_R$ index and $\alpha=\pm,\,\dot{\alpha}=\dot{\pm}$ are  Lorentz indices.}
 ${\cal Q}\equiv\bar{\cal Q}_{2+}$ $\in$ ${\cal N}=2$ superconformal algebra,
 although the index can be defined more generally.
In radial quantization the superconformal index is defined as
\be
{\cal I}= \Tr_{\cal H} (-1)^F t^{2(E+j_2)} y^{2j_1} v^{-(r+R)}\,.
\ee
The fugacities $t,y$ and $v$ couple to all possible $SU(2,2|2)$ charges that commute with ${\cal Q}$.
 $E$ is the conformal dimension. $(j_1,j_2)$ are the $SU(2)_1\otimes SU(2)_2$ Lorentz spins and $(R,r)$ are
 the charges of $SU(2)_R\times U(1)_r$ R-symmetry.
The superconformal index doesn't depend on the couplings of the theory and hence it can be calculated in the weak coupling limit.
 The entire contribution to the supersymmetric partition function on $S^3\times S^1$ thus comes from the saddle point approximation.
One loop partition function of a  $4d$ gauge theory on $S^3\times S^1$ was computed in~\cite{Aharony:2003sx} in the
 presence of fugacities associated with various conserved charges. To compute the superconformal index, we only allow
 fugacities for charges which commute with ${\cal Q}$; i.e. $t,\,y$ and $v$.

For the one loop computation in $SU(N)$ gauge theory, it is convenient to use the Coulomb gauge $\partial_i A^i=0$ where $i,j,k$
are $S^3$ coordinates and $\partial_i$ are covariant derivatives. The residual gauge freedom is fixed by imposing
 $\partial_0 \alpha=0$ where $\alpha=\frac{1}{V}\int_{S^3} A_0$ and $V$ is the volume of $S^3$.
The partition function is then written as
\be
Z=\int d\alpha \Delta_2 \int {\cal D}A \Delta_1 e^{-S(A,\alpha)}\,,
\ee
where $\Delta_1$ and $\Delta_2$ are Fadeev-Popov determinants associated with the first and second gauge fixing conditions
 respectively. For a charge $s$ that commutes with $\Q$, we can add a supersymmetric coupling with a constant background gauge field as
 \be
 S\to S+\int d^4 x\, s^\mu \chi_\mu,
 \ee
where $s^\mu$ is associated conserved current. $\chi_\mu$ is take to be a  $(\chi,0,0,0)$ and $\chi$ is identified with the chemical potential for charge $s$. The chemical potential is related to the fugacity, say $x$, of the Hamiltonian formalism as $x=e^{-\beta \chi}$. In our case, $x$ can be any of the $t,y$ and $v$.

 After performing $\int {\cal D}A$, one gets an $SU(N)$ unitary matrix model
\be
Z=\int [dU] e^{-S_{eff}[U]}\,,
\ee
where $U=e^{i\beta \alpha}$ and $\beta$ is the circumference of the circle,
$[dU]$ is the invariant Haar measure on the group $SU(N)$.
We can write $S_{eff}$ concisely as follows
\be\label{indexmodel}
S^{eff}[U]= \sum_{m=1}^\infty \frac{1}{m} \sum_j  i_{{\cal R}_j}(t^m,y^m,v^m)\chi_{{\cal R}_j} (U^m, V^m)\,.
\ee
Here, $V$ denotes the chemical potential that couples to the Cartan of the flavor group; ${\cal R}_j$ labels the representation
 of the fields under gauge and flavor groups and $i_{{\cal R}_j}$ is the single letter index of the fields in representation
 ${\cal R}_j$.

 The circumference $\beta$ of the  circle is related to the fugacity $t$ as $t=e^{-\beta/3}$.
 To produce the partition function of dimensionally reduced gauge theory on $S^3$~\cite{Kapustin:2009kz,Kapustin:2010xq}
 we also scale $v=e^{-\beta/3}$, $y=1$, and take the limit $\beta\to 0$.
In appendix \ref{app} we restore the additional deformations by defining $v=e^{-\beta(1/3+u)}$
 and set $y=e^{-\beta\eta}$ where $u$ and $\eta$ are chemical potentials for fugacities $v$ and $y$ respectively. The partition function of $3d$ gauge theories on squashed $S^3$ was computed in~\cite{Hama:2011ea}, the $\eta$ deformation is related to the squashing parameter of $S^3$.

\section{\label{mmlimit}$4d$ Index to $3d$ Partition function on $S^{3}$}

A matrix model for computing the partition function of $3d$ gauge theories on $S^3$ ($S^3$ matrix model) was obtained in \cite{Kapustin:2009kz,Kapustin:2010xq}. In this section, we will derive this matrix model as a $\beta\to0$ limit of the matrix model that computes the superconformal index (\ref{indexmodel}) (index matrix model) of the $4d$ gauge theories.
Both matrix models involve integrals over gauge group parameters and their integrand contains one-loop contributions from vector- and hyper-multiplets. We will show that the gauge group integral together with the contribution from the vector multiplet map nicely from the index model to the $S^3$ model. The contributions of the hypermultiplets match up separately. We also show that the superconformal index is the $q$-deformation of the $S^3$ partition function of the daughter theory.

\subsection{Building blocks of the matrix models}

For concreteness, let us consider $4d$ $\NN=2$ $SU(N)$ gauge theory. It is constructed using two basic building blocks: hyper-multiplets and vector multiplets.
\subsubsection*{Hyper-multiplet}
As was first observed in \cite{Dolan:2008qi}, the index of the hypermultiplet can be written elegantly in terms of a special function \cite{Gadde:2009kb}
\beq
{\cal I}^{hyp}=\prod_{i}\Gamma\left(\frac{t^{2}}{\sqrt{v}}a_i;t^{3}y,t^{3}y^{-1}\right),
\eeq
where $\Gamma$ is the elliptic gamma function \cite{Spiridonov5} defined to be
\beq
\Gamma(z;r,s)=\prod_{j,k\geq0}\frac{1-z^{-1}r^{j+1}s^{k+1}}{1-zr^{j}s^{k}}\,,
\eeq
and $a_i$ are eigenvalues of the maximal torus of the gauge/flavor group satisfying $\prod_{i=1}^N a_i=1$.
In this section, for the sake of simplicity, we set $v=t$ and $y=1$ and will discuss the general
assignment of chemical potentials in appendix~\ref{app}. We choose a convenient variable $q\equiv e^{-\beta}$ to parametrize the chemical potentials of the Cartan of the
flavor group as $a_i=q^{-i\alpha_i}$, and the chemical potential $t$ as $t=q^{\frac{1}{3}}$.
The index of the hyper-multiplet then becomes
\begin{eqnarray}
{\cal I}^{hyp}  =  \prod_{i}\prod_{j,k\geq0}\frac{1-q^{-\frac{1}{2}+i\alpha_i}q^{j+1}q^{k+1}}{1-q^{\frac{1}{2}-i\alpha_i}q^{j}q^{k}}
  =  \prod_{i}\prod_{n\geq1}\left(\frac{[n+\frac{1}{2}+i\alpha_i]_{q}}{[n-\frac{1}{2}-i\alpha_i]_{q}}\right)^{n}\,,
  \end{eqnarray}
where $[n]_{q}\equiv\frac{1-q^{n}}{1-q}$ is the  \textit{$q$-number}. It
has the property $[n]_{q}\stackrel{q\to1}{\longrightarrow}n$. So
far we have fixed the chemical potentials $v$ and $y$ that couple
to $-(R+r)$ and $j_{1}$ respectively.
To recover $3d$ partition function on $S^{3}$ we should take the
radius of $S^{1}$ to be very small, which corresponds to the limit
$q\to1$.
\begin{eqnarray}
{\cal I}^{hyp}  =  \prod_{i}\prod_{n\geq1}\left(\frac{n+\frac{1}{2}+i\alpha_i}{n-\frac{1}{2}-i\alpha_i}\right)^{n}
  =  \prod_{i}(\cosh\pi\alpha_i)^{-\frac{1}{2}}\,.
 \end{eqnarray}
One can find a proof of the second equality in~\cite{Kapustin:2009kz}. From the limiting procedure, it is clear that the superconformal index of the hypermultiplet is the $q$-deformation of the $3d$ hypermultiplet partition function.

\subsubsection*{Vector multiplet}
The index of an ${\mathcal N}=2$ vector multiplet is given by
\beq
{\cal I}^{vector}=\prod_{i< j}\frac{1}{(1-a_i/a_j)(1-a_j/a_i)}\frac{\Gamma(t^{2}v(a_i/a_j)^{\pm};t^{3}y,t^{3}y^{-1})}{\Gamma((a_i/a_j)^\pm;t^{3}y,t^{3}y^{-1})}\,,
\eeq
Here we have dropped an overall $a_i$-independent factor.
We use the condensed notation,
$\Gamma(z^{\pm1};r,s)=\Gamma(z^{-1};r,s)\Gamma(z;r,s)$. With the same variable
change as above we get
\begin{eqnarray}
{\cal I}^{vector} & = & \prod_{i<j}\frac{1}{1-q^{i(\alpha_i-\alpha_j)}}\frac{1}{1-q^{-i(\alpha_i-\alpha_j)}}\frac{1}{\Gamma(q^{\pm i(\alpha_i-\alpha_j)};q,q)}\nonumber\\
 & = & \prod_{i<j}\frac{1}{1-q^{i(\alpha_i-\alpha_j)}}\frac{1}{1-q^{-i(\alpha_i-\alpha_j)}}\prod_{n\geq1}\left(\frac{1-q^{n+i(\alpha_i-\alpha_j)+1}}{1-q^{n-i(\alpha_i-\alpha_j)-1}}\frac{1-q^{n-i(\alpha_i-\alpha_j)+1}}{1-q^{n+i(\alpha_i-\alpha_j)-1}}\right)^{-n}\\
 & \stackrel{reg}{=} & \prod_{i<j}\prod_{n\geq1}\left(\frac{[n-i(\alpha_i-\alpha_j)]_{q}}{[n]_{q}}\frac{[n+i(\alpha_i-\alpha_j)]_{q}}{[n]_{q}}\right)^{2}\,.\nonumber
 \end{eqnarray}
The last line involves regulating the infinite product in a way that
doesn't depend on $\alpha$. In the limit $q\to1$, i.e. the radius of the circle
goes to zero, we get\beq
{\cal I}^{vector}=\prod_{i<j}\prod_{n\geq1}\left(1+\frac{(\alpha_i-\alpha_j)^{2}}{n^{2}}\right)^{2}=
\prod_{i<j}\left(\frac{\sinh\pi(\alpha_i-\alpha_j)}{\pi(\alpha_i-\alpha_j)}\right)^{2}\,.\eeq
The last equality again is explained in~\cite{Kapustin:2009kz}. Again, the we see that the index of the vector multiplet is the $q$-deformation of the $3d$ vector partition function.
Most general expression for the one-loop contribution of the vector multiplet with $u$ and $\eta$ turned on is obtained in appendix \ref{app}.

\subsubsection*{Gauge group integral}

The gauge group integral in the $4d$ index matrix model is done with the invariant Haar measure
\be
[dU]=\prod_i d\alpha_i \prod_{i<j} \sin^2\(\frac{\beta(\alpha_i-\alpha_j)}{2}\)\,\,\,\stackrel{\beta\to0}{\longrightarrow}\,\,\,\prod_i d\alpha_i \prod_{i<j}\(\frac{\beta(\alpha_i-\alpha_j)}{2}\)^2.
\ee
After appropriate regularization, the measure factor precisely cancels the weight factor in the denominator of the vector multiplet one-loop determinant.
The unitary gauge group integral in the index matrix model can be done as a contour integral over $a$ variables parametrizing the Cartan sub-group, i.e. $a\in\mathbb{T}$. After the change variables $a=q^{-i\a}$ the contour integral becomes a line integral as follows.
We write  $a=q^{-i\a}=e^{i\b\a}$. The contour integral around the unit circle is then
\be
\oint_{\mathbb{T}}\frac{da}{a}\dots=\int^{\pi/\b}_{-\pi/\b}d\a\dots\qquad:\qquad
 \b\to0,\qquad \oint_{\mathbb{T}}\frac{da}{a}\dots=\int_{-\infty}^{\infty}d\a\dots\,.
\ee

\section{\label{dualities}$4d\leftrightarrow3d$ dualities}

\subsubsection*{S duality}

Let us illustrate the reduction of a four dimensional index to three
dimensional partition function with a simple example. Consider ${\mathcal N}=2$ $SU(2)$
gauge theory with four hypermultiplets in four dimensions. The index of this theory is
given by the following expression (up to overall normalization constants)
\be
\oint \frac{dz}{z}\,\frac{\Gamma(t^{3/2}a^{\pm1}b^{\pm1}z^{\pm1};t^3,t^3)
\;\Gamma(t^{3/2}c^{\pm1}d^{\pm1}z^{\pm1};t^3,t^3)}
{\Gamma(z^{\pm2};t^3,t^3)}\,.
\ee Here, $a,\,b,\,c$ and $d$ label the Cartans of $SU(2)^4\subset SO(8)$ flavor group. The Gamma
functions in the numerator come from the four hyper-multiplets; the Gamma functions in the denominator
come from the ${\mathcal N}=2$ vector multiplet.

From the results of the previous section this expression for the index
gives rise to the partition function of ${\mathcal N}=2$ $SU(2)$ gauge theory
in three dimensions. We scale $t\to 1$ and rewrite this as
 \beq
{\cal Z}(\alpha,\beta,\gamma,\delta)=\int d\sigma\frac{\sinh^{2}2\pi\sigma}{\cosh\pi(\sigma\pm\alpha\pm\beta)\cosh\pi(\sigma\pm\gamma\pm\delta)}\,,\eeq
where, each $\cosh$ is product of four factors with all sign combinations.
The flavor (now mass) parameters $\a,\,\beta,\,\gamma$ and $\delta$ are related to the flavor parameters in $4d$ as before.

The superconformal index of the ${\mathcal N}=2$ $SU(2)$
gauge theory with four hypermultiplets in four dimensions is expected to be invariant under the action of an S-duality group which
permutes the four hypermultiplets. The expression above can be explicitly shown to exhibit this property~\cite{Gadde:2009kb}.
The four dimensional S-duality implies that the three dimensional partition function is invariant under permuting $\a$, $\beta$, $\gamma$,
and $\delta$. One can show (e.g. numerically or order by order expansion in $\a$) that this is indeed true.
Note that this implies a new kind of Seiberg-like duality in three dimensions. This computation can be generalized to
any of the theories recently discussed by  Gaiotto~\cite{Gaiotto:2009we} in four dimensions. In particular the index of these theories
was claimed to posses a TQFT structure~\cite{Gadde:2009kb}; and this structure is inherited by the three dimensional partition
functions after doing the dimensional reduction. The reasoning in four dimensions and three dimensions is however different.
In four dimensions one can associate a punctured Riemann surface to each of the  superconfomal theories with the modular parameters
of the surface related to the gauge coupling constants. The index does not depend on the coupling constants and thus is independent
of the moduli giving rise to a topological quantity associated to the Riemann surface. After dimensionally reducing to  three dimensions
the theories cease to be conformal invariant and flow to a fixed point in the IR. The statement is then that at the IR fixed point
the information about the original coupling constant is ``washed away'' and theories   originally associated to  punctured Riemann
surfaces of the same topology flow to an equivalent fixed point in the IR.

\subsubsection*{Mirror symmetry}
In principle one can try to use relations special to field theories in three dimensions to gain
information about the four dimensional theories. Let us comment how this can come about.
In three dimensions certain classes of theories are related by mirror symmetry.
for example, in~\cite{Benini:2010uu} it is claimed that mirror duals of $T_{N}$~\cite{Gaiotto:2009we}
theories have Lagrangian description and are certain star shaped
quiver gauge theories. Let us see if the partition function of $T_{2}$
(free hyper-multiplet in trifundamental of $SU(2)^{3}$) matches with the partition
function of its mirror dual:
\begin{eqnarray}
{\cal Z}_{T_{2}} & = & \frac{1}{\cosh\pi(\alpha\pm\beta\pm\gamma)},\\
{\cal Z}_{\tilde{T}_{2}} & = & \int d\sigma d\mu d\nu d\rho\frac{\sinh^{2}2\pi\sigma\, e^{2\pi i(\mu\alpha+\nu\beta+\rho\gamma)}}{\cosh\pi(\sigma\pm\mu)\cosh\pi(\sigma\pm\nu)\cosh\pi(\sigma\pm\rho)}.\nonumber
\end{eqnarray}
In ${\cal Z}_{T}$, the parameters $\alpha,\beta,\gamma$ appear as
masses while in ${\cal Z}_{\tilde{T}}$ they appear as FI terms.
Let us compute ${\cal Z}_{\tilde{T}_{2}}$. 
One can perform the ${\cal Z}_{\tilde{T}_{2}}$ integrations.
First we work out
\begin{equation*}
\int d\mu\frac{e^{2\pi i\alpha\mu}}{\cosh\pi(\mu\pm\sigma)}=\frac{2\sin2\pi\alpha\sigma}{\sinh\pi\alpha\sinh2\pi\sigma}\,.
\end{equation*}
Then we find that
\begin{equation}
{\cal Z}_{\tilde{T}_{2}}=\int d\sigma\frac{8\sin2\pi\alpha\sigma\sin2\pi\beta\sigma\sin2\pi\gamma\sigma}
                                               {\sinh\pi\alpha\sinh\pi\beta\sinh\pi\gamma\sinh2\pi\sigma}\\
    =\frac{1}{\cosh\pi(\alpha/2\pm\beta/2\pm\gamma/2)}\,.
\end{equation}
${\cal Z}_{\tilde{T}_{2}}$ is actually ${\cal Z}_{T_{2}}$ if we
rescale $\alpha$, $\beta$ and $\gamma$ in ${\cal Z}_{\tilde{T}_{2}}$
by a factor of $2$. This fact can be in principle use to investigate the index of the strongly coupled SCFTs in four dimensions
which do not have Lagrangian description. One can dimensionally reduce these theories to three dimensions, consider
their mirror dual and compute its $3d$ partition function; finally, one can try to uplift this result to $4d$ and
obtain thus the superconformal index of the original four dimensional theory. The feasibility of this approach is currently
under
investigation.

\section*{Acknowledgements}
We thank Leonardo Rastelli and Shlomo Razamat for very useful discussions and guidance. We also thank Chris Beem for useful conversations.
This work was supported in part by DOE grant DEFG-0292-ER40697 and by NSF grant PHY-0969739. Any opinions, findings, and conclusions or recommendations expressed in this material are those of the authors and do not necessarily reflect the views of the National Science Foundation.
\appendix

\section{Refinement of $3d$ partition function}\label{app}
\label{sec:refinement}

The superconformal index defined in section \ref{pathint} is a function of fugacities $t,y$ and $v$. In order to recover the matrix model of Kapustin et al. \cite{Kapustin:2009kz,Kapustin:2010xq} in section \ref{mmlimit} we simply fixed the $v\rightarrow t$ and $y\rightarrow1$. In this appendix we refine the $3d$ partition function by keeping track of all the fugacities in the index. It is convenient to define the chemical potentials
\begin{equation}
  v=e^{-\beta(1/3+u)},\qquad y=e^{-\beta\eta}.
\end{equation}
The index, in terms of $\beta, u$ and $\eta$ becomes
\begin{equation}
 \II=\mathrm{Tr}(-1)^Fe^{-\beta[\frac{2}{3}(E+j_2)-\frac{1}{3}(r+R)-(r+R)u+2j_1\eta]}.
\end{equation}
Let us compute the partition function of the hypermultiplet after turning on only $u$.
\begin{equation}
  \begin{split}
    \mathcal{I}^{hyp}&=\prod_{i}\Gamma\left(\frac{t^{2}}{\sqrt{v}}a_i;t^{3}y,t^{3}y^{-1}\right)
                      =\prod_i\prod_{n\geqslant1}\left(\frac{[n+\frac{1}{2}+\frac{u}{2}+i\alpha_i]_q}{[n-\frac{1}{2}-\frac{u}{2}-i\alpha_i]_q}\right)^n\\
               &\stackrel{q\to1}{\longrightarrow}\prod_i\left[\cosh\pi\left(\alpha_i-i\frac{u}{2}\right)\right]^{-\frac{1}{2}}\\
    \mathcal{I}^{vector}&=\prod_{i<j}\frac{1}{1-q^{-i(\alpha_i-\alpha_j)}}\frac{1}{1-q^{i(\alpha_i-\alpha_j)}}
    \frac{\Gamma(q^{1+u\pm i(\alpha_i-\alpha_j)};q,q)}{\Gamma(q^{\pm
    i(\alpha_i-\alpha_j)};q,q)}\\
               &\stackrel{q\to1}{\longrightarrow}\prod_{i<j}\left(\frac{\sinh\pi(\alpha_i-\alpha_j)}{\pi(\alpha_i-\alpha_j)}\right)^{2}
  \left(\frac{\cosh\pi(\mp(\alpha_i-\alpha_j)+i/2)}{\cosh\pi(\mp(\alpha_i-\alpha_j)+i(u+1/2))}\right)^{1/2}  .
  \end{split}
\end{equation}
Both partition functions reduce to the ones in section \ref{mmlimit} as we set $u$  to zero.

Now we restore $y=q^{-\beta\eta}$ to produce the more refined $3d$ partition function. The chemical potential $\eta$ has a nice physical interpretation as the $U(1)\times U(1)$ isometry preserving squashing deformation of the $S^3$. The partition function of $3d$ gauge theories on this squashed background was computed in \cite{Hama:2011ea}.

The contribution due to the hypermultiplet with $\eta$ deformation turned on is
\begin{equation}
  \begin{split}
    \mathcal{I}^{hyp}&=\prod_i\Gamma(\frac{t^2}{\sqrt{v}}a_i;t^3y,t^3/y)\\
               &\stackrel{y\to q^{-\beta\eta}}{\longrightarrow}\prod_i\Gamma(q^{1/2-u/2-i\alpha_i};q^{1+\eta},q^{1-\eta})\\
                              &=\prod_{i}\prod_{j,k\geqslant0}
                          \frac{1-q^{3/2+u/2+i\alpha_i}q^{(1+\eta)j}q^{(1-\eta)k}}
                               {1-q^{1/2-u/2-i\alpha_i}q^{(1+\eta)j}q^{(1-\eta)k}}.
  \end{split}
\end{equation}
Using the regularized infinite product representation of Barnes' double-Gamma function
\begin{equation}
  \Gamma_2(x|\epsilon_1,\epsilon_2)\propto\prod_{m,n\geqslant0}(x+m\epsilon_1+n\epsilon_2)^{-1},
\end{equation}
the partition function of hyper-multiplet can be written in a compact way
\begin{equation}
  \begin{split}
    \mathcal{I}^{hyper}&\rightarrow\prod_i\frac{\Gamma_2(1/2-u/2-i\alpha_i|1+\eta,1-\eta)}
                                        {\Gamma_2(3/2+u/2+i\alpha_i|1+\eta,1-\eta)}\\
                  &=\prod_i\frac{\Gamma_2(\frac{Q}{2}(1/2-u/2)-i\hat\alpha_i|b,b^{-1})}
                                        {\Gamma_2(\frac{Q}{2}(3/2+u/2)+i\hat\alpha_i|b,b^{-1})},
  \end{split}
\end{equation}
where we have defined\footnote{We thank Davide Gaiotto for pointing out this change of variables.}
\be
\hat\alpha_i=\frac{\alpha_i}{\sqrt{1-\eta^2}},\qquad b=\sqrt{\frac{1-\eta}{1+\eta}},\qquad Q=b+b^{-1}.
\ee
With this change of variables it is easy to see that for $u=0$, our result is in agreement with \cite{Hama:2011ea}.
The partition function of the vector multiplet:
\begin{equation}
  \begin{split}
    \mathcal{I}^{vector}\rightarrow
    &\prod_{i<j}\frac{1}{1-q^{-i(\alpha_i-\alpha_j)}}\frac{1}{1-q^{i(\alpha_i-\alpha_j)}}
    \frac{\Gamma(q^{1+u\pm i(\alpha_i-\alpha_j)};q^{1+\eta},q^{1-\eta})}
         {\Gamma(q^{\pm i(\alpha_i-\alpha_j)};q^{1+\eta},q^{1-\eta})}.
  \end{split}
\end{equation}
reduces to
\begin{equation}
  \begin{split}
    \mathcal{I}^{vector}=
    &\prod_{i<j}\frac{(1-\eta^2)\sinh\frac{\pi(\alpha_i-\alpha_j)}{1+\eta}\sinh\frac{\pi(\alpha_i-\alpha_j)}{1-\eta}}
                                {\pi^2(\alpha_i-\alpha_j)^2}
                  \frac{\Gamma_2(1+u\pm i(\alpha_i-\alpha_j)|1+\eta,1-\eta)}
                  {\Gamma_2(1-u\pm i(\alpha_i-\alpha_j)|1+\eta,1-\eta)}\\
                  =&\prod_{i<j}\frac{\sinh\pi b (\hat\alpha_i-\hat\alpha_j)\sinh\pi b^{-1} (\hat\alpha_i-\hat\alpha_j)}
                                {\pi^2(\hat\alpha_i-\hat\alpha_j)^2}
                  \frac{\Gamma_2(\frac{Q}{2}(1+u)\pm i(\hat\alpha_i-\hat\alpha_j)|b,b^{-1})}
                  {\Gamma_2(\frac{Q}{2}(1-u)\pm i(\hat\alpha_i-\hat\alpha_j)|b,b^{-1})}.
  \end{split}
\end{equation}
Again, we find a precise agreement with the partition function of the vector multiplet on squashed $S^3$.

\bibliography{sduality}
\bibliographystyle{JHEP}
\end{document}